\documentclass[aps,pra,showpacs,showkeys,twocolumn,groupedaddress]{revtex4-1}

\usepackage{graphicx,tabularx,times,hyperref}

\setcitestyle{numbers,square}
\usepackage{array}
\usepackage{amsmath}
\usepackage{units}
\usepackage{graphicx}
\usepackage{braket}
\usepackage{xfrac}
\usepackage{enumitem}
\usepackage[normalem]{ulem}
\usepackage{booktabs}
\usepackage{tabularx}
\usepackage{multirow, bigstrut}
\usepackage{hhline}
\usepackage{times}

\usepackage{hyperref}

\hypersetup{
    colorlinks=true,
    linkcolor=cyan,
    filecolor=magenta,      
    urlcolor=blue,
}

\begin{document}


\title{Developing and evaluating an interactive tutorial on degenerate perturbation theory}
\author{Christof Keebaugh}
\author{Emily Marshman}
\author{Chandralekha Singh}
\affiliation{Department of Physics and Astronomy, University of Pittsburgh, Pittsburgh, PA 15260}
 

\begin{abstract}
 We discuss an investigation of student difficulties with degenerate perturbation theory (DPT) carried out in advanced quantum mechanics courses by administering free-response and multiple-choice questions and conducting individual interviews with students. We find that students share many common difficulties related to this topic. We used the difficulties found via research as resources to develop and evaluate a Quantum Interactive Learning Tutorial (QuILT) which strives to help students develop a functional understanding of DPT. We discuss the development of the DPT QuILT and its preliminary evaluation in the undergraduate and graduate courses. 
\end{abstract}

\maketitle

\section{Introduction}
\vspace*{-.09in}

Quantum mechanics (QM) is a particularly challenging subject for upper-level undergraduate and graduate students in physics \cite{singh, wittmann,zollman,singh2}.  
Guided by research studies conducted to identify student difficulties with QM and findings of cognitive research, we have been developing a set of research-based learning tools including the Quantum Interactive Learning Tutorials (QuILTs) \cite{zhu,singh3,singh4}.  Here, we discuss an investigation of student difficulties with degenerate perturbation theory (DPT) and the development and evaluation of a research-based QuILT that makes use of student difficulties as resources to help them develop a solid grasp of DPT.

Perturbation theory (PT) is a powerful approximate method for finding the energies and the energy eigenstates for a system for which the Time-Independent Schr\"{o}dinger Equation (TISE) is not exactly solvable.  The Hamiltonian $\hat{H}$ for the system can be expressed as the sum of two terms, the unperturbed Hamiltonian $\hat{H}^0$ and the perturbation $\hat{H}'$, i.e., $\hat{H}=\hat{H}^0+\epsilon\hat{H}'$ with $\epsilon \ll 1$. The TISE for the unperturbed Hamiltonian, $\hat{H}^0\psi_n^0 = E_n^0\psi_n^0$, is exactly solvable where $\psi_n^0$ is the $n^{th}$ unperturbed energy eigenstate and $E_n^0$ the unperturbed energy.  PT builds on the solutions of the TISE for the unperturbed case.  Using PT, the energies can be approximated as $E_n =E_n^0 + E_n^1+E_n^2 + \cdots$ where $E_n^i$ for $i=1,2,3..$ are the $i^{\textrm th}$ order corrections to the $n^{\textrm th}$ energy of the system.  The energy eigenstate can be approximated as $\psi_n = \psi_n^0+\psi_n^1+\psi_n^2 + \cdots$ where $\psi_n^i$ are the 
 $i^{\textrm th}$ order corrections to the n$^{\textrm th}$ energy eigenstate.  The tutorial focuses on the following first order perturbative corrections to the energies and energy eigenstates, which are usually the dominant corrections:
\begin{equation}\label{energy}
E_n^1 = \langle \psi_n^0|\hat{H}'|\psi_n^0\rangle,\,\,\,
\vspace*{-.16in}
|\psi_n^1\rangle  = \sum_{m \neq n} \frac{\langle \psi_m^0|\hat{H}'|\psi_n^0\rangle }{(E_n^0-E_m^0)}|\psi_m^0\rangle .
\end{equation}
In  Eq. \ref{energy}, $\left\{|\psi_n^0\rangle \right\}$ is a complete set of eigenstates of 
$\hat{H}^0$. 
When the eigenvalue spectrum of $\hat{H}^0$ has degeneracy (two or more eigenstates of $\hat{H}^0$ have the same energy, i.e., two or more diagonal elements of $\hat{H}^0$ are equal), Eq. \ref{energy} from nondegenerate perturbation theory (NDPT) is still valid provided one uses a {\it good} basis.   
For a given $\hat{H}^0$ and $\hat{H}'$, we define a {\it good} basis as consisting of a complete set of eigenstates of $\hat{H}^0$  that diagonalizes $\hat{H}'$ in each degenerate subspace of $\hat{H}^0$ and keeps $\hat{H}^0$ diagonal. In a {\it good} basis, $\hat{H}'$ is diagonal in each degenerate subspace of $\hat{H}^0$. Therefore, the terms $\langle \psi_m^0|\hat{H}'|\psi_n^0\rangle$ in Eq. \ref{energy} for the wavefunction are zero when $m \neq n$ so that the expression for the corrections to the wavefunction in Eq. \ref{energy} do not have terms that diverge.  In a {\it good} basis, Eq. \ref{energy} is also valid for finding the first order corrections to the energies (which are the diagonal elements of the $\hat{H}'$ matrix as given by Eq. \ref{energy}).

Degenerate perturbation theory is challenging for students since not only does it require an understanding of QM but also requires a strong background in linear algebra.  Students should understand that when the originally chosen basis is not a {\it good} basis for a given $\hat{H}^0$ and $\hat{H}'$, but consists of a complete set of eigenstates of $\hat{H}^0$, a {\it good} basis can be constructed from the originally chosen basis states by diagonalizing the $\hat{H}'$ matrix in each degenerate subspace of $\hat{H}^0$. Students must also be able to identify degenerate subspaces of $\hat{H}^0$, understand why both $\hat{H}^0$ and $\hat{H}'$ should be examined carefully to determine if the original basis is {\it good} and that they should only diagonalize the $\hat{H}'$ matrix in each degenerate subspace of $\hat{H}^0$ (instead of diagonalizing the entire $\hat{H}'$ matrix) to find a {\it good} basis if the original basis is not {\it good}. They should also understand
 why such a basis transformation does not change the diagonal nature of $\hat{H}^0$ (it is essential that the basis states are eigenstates of $\hat{H}^0$ in PT because the corrections to the unperturbed energies are small). After all of these considerations, students can use Eq. \ref{energy} to determine the perturbative corrections.   

\vspace*{-.19in}

\section{Student Difficulties}
\vspace*{-.11in}

Student difficulties with DPT were investigated using responses from 32 upper-level undergraduate students' to open-ended and multiple-choice questions administered after traditional instruction in relevant concepts and responses from 10 students' during individual think-aloud interviews.  Below, we discuss three of the common student difficulties with DPT found via research in the context of a three dimensional Hilbert space with a two-fold degeneracy. In particular, we find that when students are given the Hamiltonian for a system, they have difficulty correctly (1) identifying $\hat{H}'$ in the degenerate subspace of $\hat{H}^0$, (2) identifying whether the originally chosen basis is a {\it good} basis for finding the perturbative corrections, and (3) then finding a {\it good} basis if the originally chosen basis is not already a {\it good} basis. 

{\bf A. Difficulty identifying $\hat{H}'$ in the degenerate subspace of $\hat{H}^0$ given the Hamiltonian $\hat{H}$:}  
Many students had difficulty identifying the $\hat{H}'$ matrix in the degenerate subspace of $\hat{H}^0$, when the Hamiltonian $\hat{H}$ for the system was provided in a matrix form.  In particular, many students did not understand that in order to determine $\hat{H}'$ in the degenerate subspace of $\hat{H}^0$, they should start by identifying whether there is degeneracy in the energy spectrum of $\hat{H}^0$.  In fact, we find that many students incorrectly focused on the diagonal elements of the perturbation $\hat{H}'$ to determine whether there was degeneracy in the system and whether they should use DPT.   
However, degeneracy in $\hat{H}'$ has nothing to do with whether one should use DPT and whether one should examine that the original basis, in which $\hat{H}^0$ and $\hat{H}'$ are provided, is {\it good}. 

{\bf B. Difficulty {\it identifying} whether the originally chosen basis is a {\it good} basis for finding perturbative corrections:}  A {\it good} basis is one that keeps the unperturbed Hamiltonian $\hat{H}^0$ diagonal while diagonalizing the perturbation $\hat{H}'$ in the degenerate subspace of $\hat{H}^0$.  However, several students incorrectly stated that the originally chosen basis is a {\it good} basis because it consists of a complete set of eigenstates of $\hat{H}^0$ ($\hat{H}^0$ was diagonal in the original basis) without any consideration for whether $\hat{H}^0$ had any degeneracy and the implications of the degeneracy in $\hat{H}^0$ for what should be examined in the $\hat{H}'$ matrix before using Eq. \ref{energy}.  Other students only examined the basis in a general manner and did not focus on either $\hat{H}^0$ or $\hat{H}'$.  For example, one student incorrectly stated that the basis is a {\it good} basis if ``it forms a complete Hilbert space."  Another student stated that ``the basis vectors should be orthogonal" is the only condition to have a {\it good} basis regardless of the fact that the unperturbed Hamiltonian $\hat{H}^0$ had degeneracy in the situation provided.  

Furthermore, when students were given a Hamiltonian $\hat{H}= \hat{H}^0+\hat{H}'$ in a basis (which consisted of a complete set of eigenstates of $\hat{H}^0$) and asked if that basis is a {\it good} basis, some students had a tendency to focus on either $\hat{H}^0$ or $\hat{H}'$ but not both as is necessary to correctly answer the question.  For example, during the interview, one student said, ``$\hat{H'}$ must be diagonal in the {\it good} basis". Equivalently, another student claimed the basis was not a {\it good} basis ``since $\hat{H}'$ has off-diagonal terms in this basis." These types of incorrect responses suggest that students have difficulty with the fact that $\hat{H'}$ should only be diagonal in the degenerate subspace of $\hat{H}^0$.
Students with these types of responses focused on diagonalizing the entire $\hat{H}'$ matrix (rather than diagonalizing $\hat{H}'$ in the degenerate subspace of $\hat{H}^0$). They did not realize or consider the fact that if $\hat{H}^0$ and $\hat{H}'$ do not commute, $\hat{H}^0$ may become non-diagonal if the entire $\hat{H}'$ matrix is diagonalized .  

Moreover, some students had difficulty with the fact that even when the originally chosen basis is not a {\it {\it good}} basis, it may include some states that are {\it {\it good}} states (the sub-basis in the degenerate subspace is not good but the sub-basis in the non-generate subspace is good) and can be used to find the perturbative corrections using Eq. \ref{energy}.  When students were asked to identify whether the originally chosen states were {\it {\it good}} states, roughly one-fourth 
of students 
were unable to correctly identify whether each state in the originally chosen basis is a {\it {\it good}} state or not.   For example, during the interview, one students said, ``We cannot trust nondegenerate basis states for finding corrections to the energy.   We must adjust all the basis states since we can't guarantee any will be the same."  The students with this type of response assumed that if the unperturbed Hamiltonian has degeneracy then none of the originally chosen basis states are {\it good} states. However, any state belonging to the nondegenerate subspace of $\hat{H}^0$ is a {\it {\it good}} state.  Many other students had similar difficulty.  

Also, other students struggled with the fact that if $\hat{H}'$ is already diagonal in a degenerate subspace of $\hat{H}^0$ in the original basis, 
the originally chosen basis is a {\it {\it good}} basis, and Eq. \ref{energy} can be used to determine the perturbative corrections without additional work to diagonalize $\hat{H}'$ in the subspace. They attempted to diagonalize a matrix that was already diagonal.

{\bf C. Difficulty {\it finding} a {\it good} basis if the originally chosen basis is not already a {\it good} basis for finding the perturbative corrections:} Many students struggled to find a {\it good} basis if the originally chosen basis was not already a {\it good} basis.  They did not realize that when the originally chosen basis is not already a {\it good} basis and the unperturbed Hamiltonian $\hat{H}^0$ and the perturbing Hamiltonian $\hat{H}'$ do not commute, they must diagonalize $\hat{H}'$ matrix only in the degenerate subspace of $\hat{H}^0$.  The most common mistake was diagonalizing the entire $\hat{H}'$ matrix instead of diagonalizing the $\hat{H}'$ matrix only in the degenerate subspace of $\hat{H}^0$. For example, roughly half 
of students 
diagonalized the entire $\hat{H}'$ matrix.  When asked to determine a {\it good} basis for a Hamiltonian in which $\hat{H}^0$ and $\hat{H}'$ do not commute, one interviewed student incorrectly stated, ``We must find the simultaneous eigenstates of $\hat{H}^0$ and $\hat{H}'$."  This student and many others did not realize that when $\hat{H}^0$ and $\hat{H}'$ do not commute, we cannot simultaneously diagonalize $\hat{H}^0$ and $\hat{H}'$ since they  do not share a complete set of eigenstates.  Students struggled with the fact that if $\hat{H}^0$ and $\hat{H}'$ do not commute, then diagonalizing $\hat{H}'$ produces a basis in which $\hat{H}^0$ is no longer diagonal.  Since $\hat{H}^0$ is the dominant term and $\hat{H}'$ provides only small corrections, we must ensure that the basis states used to determine the perturbative corrections in Eq. \ref{energy} remain eigenstates of $\hat{H}^0$.  If $\hat{H}^0$ and $\hat{H}'$ do commute, it is possible to diagonalize $\hat{H}^0$ and $\hat{H}'$ simultaneously to find a complete set of shared eigenstates.  However, diagonalizing $\hat{H}'$ only in the degenerate subspace of $\hat{H}^0$ still produces a {\it good} basis and in general requires less algebra and fewer opportunities for mistakes. Students had difficulty with these issues partly because they did not have a solid foundation in linear algebra in order to apply it in the context of DPT.

Some students struggled with the fact that $\hat{H}'$ should be diagonalized in the degenerate subspace of $\hat{H}^0$ while keeping $\hat{H}^0$ diagonal.  For example, one student in the interview stated, ``We cannot diagonalize a part of $\hat{H}'$, we must diagonalize the whole thing."  Students, in general, had great difficulty with the fact that the degeneracy in the eigenvalue spectrum of $\hat{H}^0$ provides the flexibility in the choice of basis in the degenerate subspace of $\hat{H}^0$, so that $\hat{H}'$ can be diagonalized in that subspace (even if $\hat{H}^0$ and $\hat{H}'$ do not commute) while keeping $\hat{H}^0$ diagonal.  For example, if we consider the case in which $\hat{H}^0$ has a two-fold degeneracy, then
\mbox{$\hat{H}^0 \psi_a^0 = E^0 \psi_a^0,  \,\,\hat{H}^0 \psi_b^0 = E^0 \psi_b^0, \,\, \textrm{and} \,\, \langle \psi_a^0|\psi_b^0 \rangle =0$}
where $\psi_a^0$ and $\psi_b^0$ are normalized degenerate eigenstates of $\hat{H}^0$.  Any linear superposition of these two states, say $\psi^0 = \alpha \psi_a^0 + \beta \psi_b^0$, with $\vert \alpha \vert^2+\vert \beta \vert^2=1$ must remain an eigenstate of $\hat{H}^0$ with the same energy $E^0$.  
Many students struggled to realize that since any linear superposition of the original basis states that correspond to the degenerate subspace of $\hat{H}^0$ remains an eigenstate of $\hat{H}^0$, we can choose that special linear superposition that diagonalizes $\hat{H}'$ in the degenerate subspace of $\hat{H}^0$.  


\vspace*{-.19in}
\section{Development of the QuILT}
\vspace*{-.11in}

The development of the DPT QuILT started with an investigation of student difficulties via open-ended and multiple-choice questions administered after traditional instruction to advanced undergraduate and graduate students and conducting a cognitive task analysis from an expert perspective of the requisite knowledge \cite{cognitive}. The QuILT strives to help students build on their prior knowledge and addresses common difficulties. After a preliminary version was developed based upon the task analysis \cite{cognitive} and knowledge of common student difficulties, it underwent many iterations among the three researchers and then was iterated several times with three physics faculty members to ensure that they agreed with the content and wording. It was also administered to advanced undergraduate students in individual think-aloud interviews to ensure that the guided approach was effective, the questions were unambiguously interpreted, and to better understand the rationale for student responses. During these semi-structured interviews, students were asked to “think aloud” while answering the questions. Students first read the questions on their own and answered them without interruptions except that they were prompted to think aloud if they were quiet for a long time. After students had finished answering a particular question to the best of their ability, they were asked to further clarify and elaborate on issues that they had not clearly addressed earlier. Modifications and improvements were made based upon the student and faculty feedback. 

The QuILT uses an inquiry-based approach to learning and actively engages students in the learning process. It includes a pretest to be administered in class after instruction in DPT. Then students engage with the tutorial in small groups in class or alone when using it as a self-paced learning tool in homework, and then they are administered a posttest in class. As students work through the tutorial, they are asked to predict what should happen in a given situation. Then, the tutorial strives to provide scaffolding and feedback as needed to bridge the gap between their initial knowledge and the level of understanding that is desired.  Students are also provided checkpoints to reflect upon what they have learned and to make explicit the connections between what they are learning and their prior knowledge. They are given an opportunity to reconcile  differences between their predictions and  the guidance provided in the checkpoints before proceeding further. 
 
In the QuILT, students actively engage with examples involving DPT that are restricted to a three dimensional Hilbert space (with two-fold degeneracy) to allow them to focus on the fundamental concepts without requiring cumbersome calculations that may detract from the focus on why it is important to determine if the original basis is a {\it good} basis, and if it is not {\it good}, what type of basis transformation must be performed before Eq. \ref{energy}  can be used to find the corrections.  In particular, for a given $\hat{H}^0$ and $\hat{H}'$, when there is degeneracy in the eigenvalue spectrum of $\hat{H}^0$, students learn about why all bases are not {\it good} even though they may consist of a complete set of eigenstates of $\hat{H}^0$,  how to determine if the basis is {\it good} and how to change the basis to one which is {\it good} (if the original basis is not {\it good}) so that Eq. \ref{energy} can be used to find the first order corrections. 

In the QuILT, students work through different examples in which the same unperturbed Hamiltonian $\hat{H}^0$ is provided and they are asked to identify whether the originally chosen basis is a {\it good} basis for a given $\hat{H}'$.  In one example, $\hat{H}'$ is diagonal in the degenerate subspace of $\hat{H}^0$ in the original basis provided and therefore the original basis is a {\it good} basis.  In another example, 
the perturbation $\hat{H}'$ is not diagonal in the degenerate subspace of $\hat{H}^0$ and therefore the original basis is not a {\it good} basis.  
These examples help students learn that consideration of both $\hat{H}^0$ and $\hat{H}'$ is necessary to determine if the basis is {\it good} in DPT. Students are then asked to summarize in their own words why the original basis is a {\it good} basis or not.

\vspace*{-.19in}
\section{Preliminary Evaluation}
\vspace*{-.11in}

Once the researchers determined that the QuILT was successful in one-on-one implementation using a think-aloud protocol, it was given to 11 upper-level undergraduates in a second-semester junior/senior level QM course and 19 first-year physics graduate students in the second-semester of the graduate core QM course.  Both undergraduate and graduate students were given a pretest after traditional instruction in relevant concepts in DPT but before working through the tutorial.  The undergraduates worked through the tutorial in class for two days and were asked to work on the remainder of the tutorial as homework.  The graduate students were given the tutorial as their weekly homework assignment.  After working through and submitting the completed tutorial, both groups were given the posttest with questions similar to the pretest but with the degenerate subspace of $\hat{H}^0$ being different.  The following were the pretest questions:

1. Consider the unperturbed Hamiltonian \\
$\hat{H}^0 =  V_0
\left[
\begin{array}{rrr}
3&0&0\\
0&3&0\\
0&0&7
\end{array}
\right].$\\
(a)  Write an example of a perturbing Hamiltonian $\hat{H}'$ in the same basis as $\hat{H}^0$ such that for that $\hat{H}^0$ and $\hat{H}'$, this basis forms a {\it good} basis (so that one can use the same expressions that one uses in non-DPT for perturbative corrections). 
Use $\epsilon$ as a small parameter.


(b)  Write an example of a perturbing Hamiltonian $\hat{H}'$ in the same basis as $\hat{H}^0$ such that for that $\hat{H}^0$ and $\hat{H}'$, this basis does NOT form a {\it good} basis 
(so that we cannot use the basis for perturbative corrections using Eq. \ref{energy}).

Use $\epsilon$ as a small parameter.

2. Given $\hat{H} = \hat{H}^0 + \epsilon\hat{H}' = V_0
\left[
\begin{array}{rrr}
5&0&-4\epsilon\\
0&1-4\epsilon&0\\
-4\epsilon&0&1+6\epsilon
\end{array}
\right]$ with $\epsilon \ll 1$, determine the first order corrections to the energies.  {\bf You must show your work.}

3. Given $\hat{H} = \hat{H}^0 + \epsilon\hat{H}' = V_0
\left[
\begin{array}{rrr}
2&\epsilon&\epsilon\\
\epsilon&2&\epsilon\\
\epsilon&\epsilon&3
\end{array}
\right]$, with\\ $\epsilon \ll 1$, determine the first order corrections to the energies.  {\bf You must show your work.}

\begin{table}[!tbp]
  \caption[Comparison Group vs Incentivized Group Gains]{Average pretest and posttest scores, gain ($G$) and normalized gain ($g$) for undergraduate students (number of students $N=11$) and graduate students (number of students $N=19$). }
  \label{prepostpercent}
  \centering
  \begin{tabularx}{\linewidth}{>{\itshape}l XXXX c XXXX}
 
 \toprule
  \hline
  & \multicolumn{4}{c}{Undergraduate Students} & &\multicolumn{4}{c}{Graduate Students} \\
  \hhline{~----~----}

  & Pre & Post &\centering G&\centering g& & Pre & Post & \centering G& \hspace*{.07in} g \\
  \hhline{-----~----} 
   \multicolumn{1}{l}{1(a)}   &{ 23.1} &{ 100} &{ +76.9}& { 1.00}  & &{ 67.5} &{ 88.2} &{ +20.7} & { 0.64}\\
   {1(b)} &15.4 &100 &+84.6& 1.00&   &51.3 &73.7 &+22.4&0.46 \\
  2  &32.3 &100 &+67.7&1.00&    &30.0 &94.8 &+64.8&0.93 \\
  3  &2.6 &100 &+97.4& 1.00&   &11.7 &86.0 &+74.3&0.84 \\
  \hline
  
  \bottomrule
  \end{tabularx}
\end{table}

Question 1 
focuses on student difficulties A and B.  In Question 1, students must be able to correctly identify the degenerate subspace of $\hat{H}^0$.  They must then determine how to construct an $\hat{H}'$ matrix in order to make sure that the basis used to represent $\hat{H}^0$ and $\hat{H}'$ in matrix form is a {\it good} basis in part (a) and not a {\it good} basis in part (b).  For question 1(a), in order for the basis to be a {\it good} basis, the constructed $\hat{H}'$ matrix must be diagonal in the degenerate subspace of $\hat{H}^0$.  For question 1(b), in order for the basis not to be a {\it good} basis, the $\hat{H}'$ matrix must be non-diagonal in the degenerate subspace of $\hat{H}^0$.


Question 2 also focuses  on student difficulties A and B.  Students must first identify $\hat{H}'$ and $\hat{H}^0$ in the degenerate subspace of $\hat{H}^0$.  Once they identify $\hat{H}'$ in the degenerate subspace of $\hat{H}^0$, they must determine whether the originally chosen basis is a {\it good} basis.  In particular, they must realize that in question 2, $\hat{H}'$ is diagonal in the degenerate subspace of $\hat{H}^0$ and therefore the original basis is a {\it good} basis.  

Question 3 focuses  on student difficulties A, B, and C.  Students must first identify $\hat{H}'$ and $\hat{H}^0$ in the degenerate subspace of $\hat{H}^0$.  Once they identify $\hat{H}'$ in the degenerate subspace of $\hat{H}^0$, they must determine whether the originally chosen basis is a {\it good} basis.  In question 3, $\hat{H}'$ is not diagonal in the degenerate subspace of $\hat{H}^0$. Thus, the original basis is not a {\it good} basis and they must find a {\it good} basis for perturbative corrections.  

The open-ended questions were graded using rubrics which were developed by the researchers together. A subset of questions was graded separately by them. After comparing 
the grading, they discussed any disagreements and resolved them 
with a final inter-rater reliability of better than 90\%. Table \ref{prepostpercent} shows the performance of undergraduates and graduate students on the pretest and posttest.  The pretest was scored for completeness for both groups but the posttest counted differently towards the course grade for the two groups.  One reason for why the undergraduates' performance on the posttest is better than that of graduate students may be that the course grade for the posttest was based on correctness for the undergraduates but on completeness for the graduate students. Table \ref{prepostpercent} also includes the average gain, $G$, and normalized gain~\cite{hake}, $g$.  The normalized gain is defined as the posttest percent minus the pretest percent divided by (100-pretest percent). The posttest scores are significantly better than the pretest scores on all of these questions for both groups.

To investigate retention of learning, the undergraduates were given questions 1(a) and 1(b) again as part of their final exam.  The final exam was six weeks after students engaged with the tutorial.   
The average score on question 1(a) was 97.8\% and on question 1(b) was 91.0\%.  In question 1(a), all 11 students provided an $\hat{H}'$ matrix that was diagonal in the degenerate subspace of $\hat{H}^0$.
In question 1(b), 10 out of 11 students provided an $\hat{H}'$ matrix that was not diagonal in the degenerate subspace of $\hat{H}^0$.
These results 
are encouraging.

\vspace*{-.2in}
\section{Summary}
\vspace*{-.1in}
Using the common difficulties of advanced students with DPT as resources, we developed and evaluated a research-based QuILT which focuses on helping students reason about and find the perturbative corrections using DPT.
It strives to provide appropriate scaffolding and feedback using a guided inquiry-based approach to help students develop a functional understanding of DPT.  The preliminary evaluation shows that the QuILT was effective in improving undergraduate and graduate students' understanding of the fundamentals of DPT.  

\vspace*{-.2in}
\section*{Acknowledgements}
\vspace*{-.1in}
We thank the NSF for award PHY-1505460.
\vspace*{-.27in}

\end{document}